\documentclass[apjl]{emulateapj}

\usepackage{color}

\shorttitle{First high-resolution images of the Sun in the $2796$\,\AA{} M\lowercase{g}\,{\sc ii}\,\lowercase{k} line}
\shortauthors{Riethm\"uller et al.}

\begin{document}

\renewcommand{\thefootnote}{\fnsymbol{footnote}}
\renewcommand{\thempfootnote}{\fnsymbol{mpfootnote}}

\newcommand{\sunrise}{\textsc{Sunrise}}
\newcommand{\carcsec}{$\mbox{.\hspace{-0.5ex}}^{\prime\prime}$}

\title{First high-resolution images of the Sun in the $2796$\,\AA{} M\lowercase{g}\,{\sc ii}\,\lowercase{k} line}

\author{\textsc{
T.~L.~Riethm\"uller,$^{1}$
S.~K.~Solanki,$^{1,2}$
J.~Hirzberger,$^{1}$
S.~Danilovic,$^{1}$
P.~Barthol,$^{1}$
T.~Berkefeld,$^{3}$
A.~Gandorfer,$^{1}$
L.~Gizon,$^{1,4}$
M.~Kn\"olker,$^{5}$
W.~Schmidt,$^{3}$
\& J.~C.~Del~Toro~Iniesta$^{6}$
}}
\affil{
$^{1}$Max-Planck-Institut f\"ur Sonnensystemforschung, Max-Planck-Str. 2, 37191 Katlenburg-Lindau, Germany; riethmueller@mps.mpg.de\\
$^{2}$School of Space Research, Kyung Hee University, Yongin, Gyeonggi, 446-701, Republic of Korea\\
$^{3}$Kiepenheuer-Institut f\"ur Sonnenphysik, Sch\"oneckstr. 6, 79104 Freiburg, Germany\\
$^{4}$Institut f\"ur Astrophysik Georg-August-Universit\"at G\"ottingen, Friedrich-Hund-Platz 1, 37077 G\"ottingen, Germany\\
$^{5}$High Altitude Observatory, National Center for Atmospheric Research, P.O. Box 3000, Boulder, CO 80307-3000, USA\\
$^{6}$Instituto de Astrof\'{\i}sica de Andaluc\'{\i}a (CSIC), Apartado de Correos 3004, 18080 Granada, Spain\\
}

\begin{abstract}
   We present the first high-resolution solar images in the Mg\,{\sc ii}\,k $2796$\,\AA{} line.
   The images, taken through a $4.8$\,\AA{} broad interference filter, were obtained during the second science
   flight of \sunrise{} in June 2013 by the SuFI instrument. The Mg\,{\sc ii}\,k images display
   structures that look qualitatively very similar to images taken in the core of Ca\,{\sc ii}\,H.
   The Mg\,{\sc ii} images exhibit reversed granulation (or shock waves) in the internetwork regions
   of the quiet Sun, at intensity contrasts that are similar to those found in Ca\,{\sc ii}\,H. Very prominent
   in Mg\,{\sc ii} are bright points, both in the quiet Sun and in plage regions, particularly near
   disk center. These are much brighter than at other wavelengths sampled at similar resolution. Furthermore,
   Mg\,{\sc ii}\,k images also show fibril structures associated with plage regions. Again, the
   fibrils are similar to those seen in Ca\,{\sc ii}\,H images, but tend to be more pronounced,
   particularly in weak plage.   
\end{abstract}


\section{Introduction}

   A good knowledge and understanding of the chromosphere
   is essential for making progress in the questions of chromospheric and coronal heating. Studying the
   chromosphere is made challenging by the fact that only few spectral lines in the visible and IR spectral
   ranges accessible from the ground are formed there. Of these even fewer sample the middle and upper
   chromosphere (He\,{\sc i} 10830, H$\alpha$, Ca\,{\sc ii} lines). Therefore, it is of considerable interest
   to explore further avenues. Besides lines in the vacuum ultraviolet, e.g. the Ly alpha and neutral carbon and oxygen lines,
   observed by e.g. SUMER \citep{Wilhelm1995,Judge1997,Carlsson1997} or VAULT \citep{Vourlidas2010}, another
   source of information on the chromosphere are the h and k lines of Mg\,{\sc ii}. These lines are expected
   to sample a large height range, reaching from the lower to the upper chromosphere, but observations made
   in these lines are rare due to their inaccessibility from ground. In the mid-1970s,
   space-based observations of Mg\,{\sc ii} h\&k spectra were carried out with the LPSP instrument
   onboard the OSO-8 mission \citep{Bonnet1978,Kneer1981}, followed in the early 1980s by observations with
   the Ultraviolet Spectrometer and Polarimeter (UVSP) onboard the Solar Maximum Mission \citep{Woodgate1980}.
   A somewhat higher spatial resolution (one or more arcsec) was reached with the 30~cm telescopes of the
   balloon-borne spectrograph RASOLBA \citep{Staath1995} in 1986 as well as with the sounding-rocket
   experiments HRTS-9 in 1995 \citep{Morrill2001,Morrill2008} and the Solar Ultraviolet
   Magnetograph Investigation (SUMI) in 2010 \citep{West2007,West2011}. Mg\,{\sc ii} h\&k spectra and slit
   jaw images are also being recorded with the 20~cm telescope of the Interface Region Imaging Spectrograph
   \citep[IRIS;][]{Wuelser2012}, which was launched on 2013 June 27, i.e. after the images discussed here
   were taken. 

   In this paper we present the first high-resolution Mg\,{\sc ii}\,k images of quiet and active solar
   regions which were recorded with the Sunrise Filter Imager \citep[SuFI;][]{Gandorfer2011} during the
   second science flight of the balloon-borne observatory \sunrise{} \citep{Solanki2010,Barthol2011} in
   June 2013. Its 1~m primary mirror makes \sunrise{} the largest solar telescope to have left Earth's surface.

\begin{figure}
   \centering
   \includegraphics[width=\linewidth]{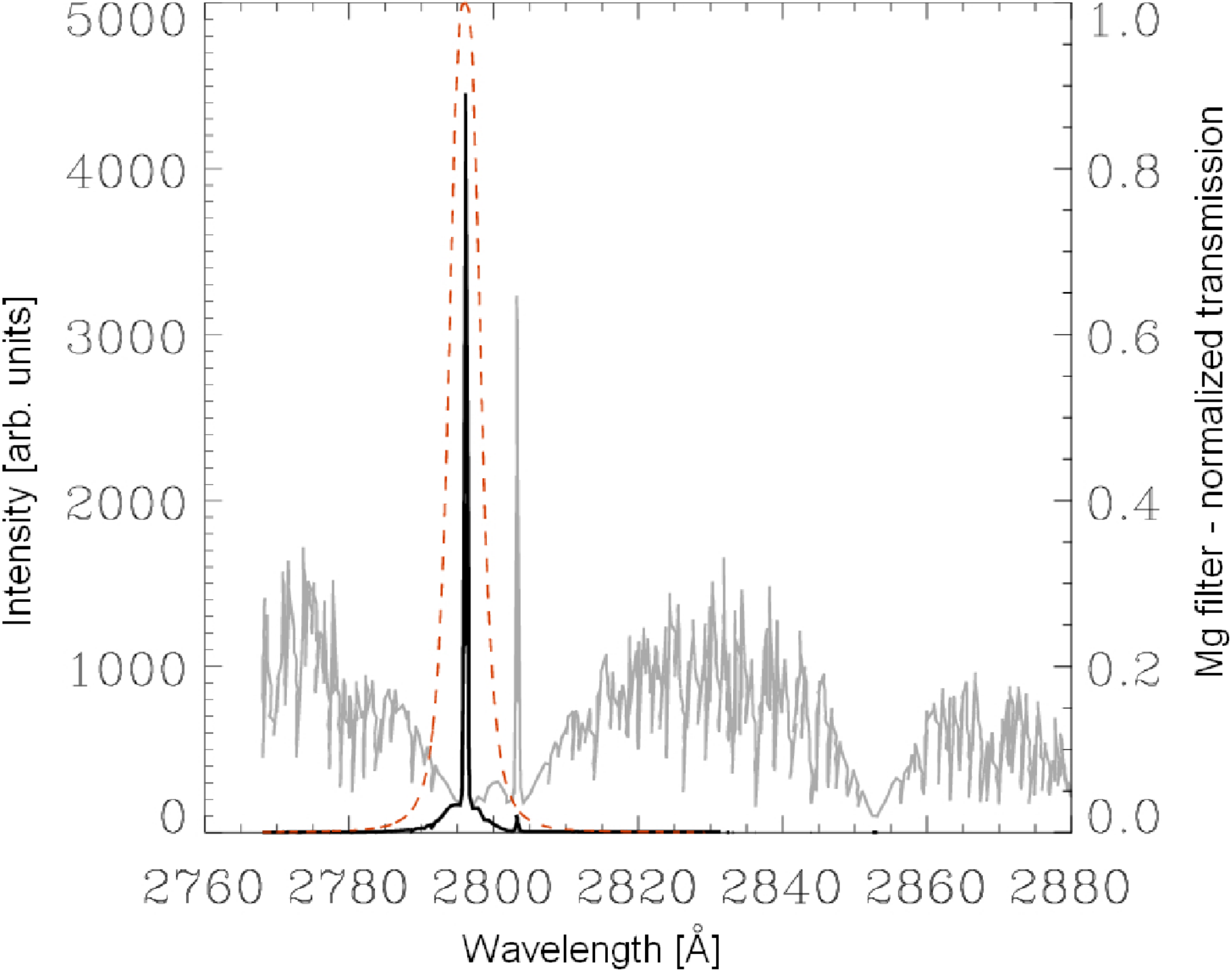}
   \caption{Observed Mg\,{\sc ii} h\&k spectrum in plage (solid grey) and the same multiplied with the
   SuFI filter transmission curve (solid black; multiplied by 10 for better display). The SuFI Mg filter
   transmission profile is overplotted with corresponding y axis on the right (dashed red).}
   \label{Fig1}
   \end{figure}

\section{Observations and data reduction}

   The second science flight of the ballon-borne solar observatory \sunrise{} started on 2013 June 12 from
   ESRANGE near Kiruna in northern Sweden. It reached a mean float altitude of $37-38$\,km, allowing seeing-free
   observations in the visible and near-ultraviolet spectral range. After a total flight time of 127 hours,
   the balloon landed safely in Boothia, a peninsula in northern Canada. Technical details of the 1~m aperture
   Gregory telescope are described by \citet{Barthol2011}. Image stabilization and feature tracking were
   achieved by the gondola's pointing system in conjunction with a six-element Shack$-$Hartmann correlating
   wavefront sensor (CWS) which controlled the telescope's focus mechanism, and a tip/tilt mirror \citep{Berkefeld2011}.

   The observations presented here were taken by the Sunrise Filter Imager \citep[SuFI;][]{Gandorfer2011}.
   Some of the filters from the 2009 flight, described by \cite{Gandorfer2011}, were replaced to put more
   emphasis on the chromosphere. Of greatest interest for the purposes of the present paper is the addition
   of a filter with a full-width-half-maximum (FWHM) of $4.8$\,\AA{} centered on $2796$\,\AA{}. Figure~\ref{Fig1}
   displays the filter profile overplotted on the solar spectrum \citep[average plage;][]{Morrill2001}.
   The filter covers the core of the Mg\,{\sc ii}\,k line, but also gets a minor contribution
   from Mg\,{\sc ii}\,h and part of the wings of the two lines, as well as some weak photospheric absorption features.
   Due to the strength of the emission core, this provides much of the signal in the filter, although
   the wings also contribute. Note that in the quiet Sun the contribution of photospheric features
   (reversed granulation) dominates for broad-band filters \citep[e.g.][]{Reardon2009,Beck2013}. In plage regions
   the chromospheric contribution is enhanced due to the increased strength of the Mg\,{\sc ii} h\&k emission peaks.

   Another important change was the inclusion of a second wavelength band centered on the core of the
   Ca\,{\sc ii}\,H $3968$\,\AA{}, in addition to the $1.8$\,\AA{} wide filter that already flew on the 2009 flight.
   The new $1.1$\,\AA{} wide filter gets more contribution from the H$_2$ and H$_3$ features. 

   Here we present first results from the Mg\,{\sc ii}\,k channel. Further descriptions of
   instrumentation updates relative to the 2009 flight as well as of the flight and the full data set obtained will
   be given elsewhere. 

   The Mg\,{\sc ii}\,k line lies in the Hartley absorption band of ozone, close to the
   maximum of absorption around $2550$\,\AA{}. Even at the float altitude of \sunrise{},
   there is sufficient stratospheric ozone to significantly attenuate the Sunlight reaching the
   telescope \citep[we estimate roughly by $2-3$ orders of magnitude, based on the modelling
   by][]{Solomon2012}. To reduce the influence of this absorption, we restricted observations
   of Mg\,{\sc ii}\,k to times around local noon, when the solar elevation was above $\mathrm{30^\circ}$,
   and scheduled ballast drops to the morning hours to ensure that the payload was at maximum
   altitude around local noon. Since \sunrise{}'s flight trajectory lay
   to the north of the arctic circle, the Sun's elevation never exceeded about $\mathrm{45^\circ}$.
   Even after these measures, the Mg\,{\sc ii} images needed to be integrated for $50$\,s and
   the S/N rarely exceeded 25-30 in the quiet Sun. 
  
   We concentrate here on the first presentation of the Mg\,{\sc ii} images and compare them with SuFI's
   Ca\,{\sc ii}\,H and $3000$\,\AA{} observations. The $3000$\,\AA{} filter was unchanged compared to the 2009
   flight and we refer to \cite{Gandorfer2011} for further details. 

   SuFI is equipped with a phase-diversity (PD) prism. After the images were dark-current
   corrected and flat-fielded, they were PD reconstructed using individual wavefront errors
   \citep[level-2 data, see][for more details]{Hirzberger2010,Hirzberger2011}.
   All intensity images were normalized to the intensity level of the mean quiet Sun, $I_{\rm{QS}}$,
   defined as the average of the whole image. SuFI's plate scale is 0\carcsec{}0207~pixel$^{-1}$ and the
   field of view (FOV) was about 15\arcsec{}$\times$39\arcsec{}.

\section{Results}

\begin{figure}
   \centering
   \includegraphics[width=\linewidth]{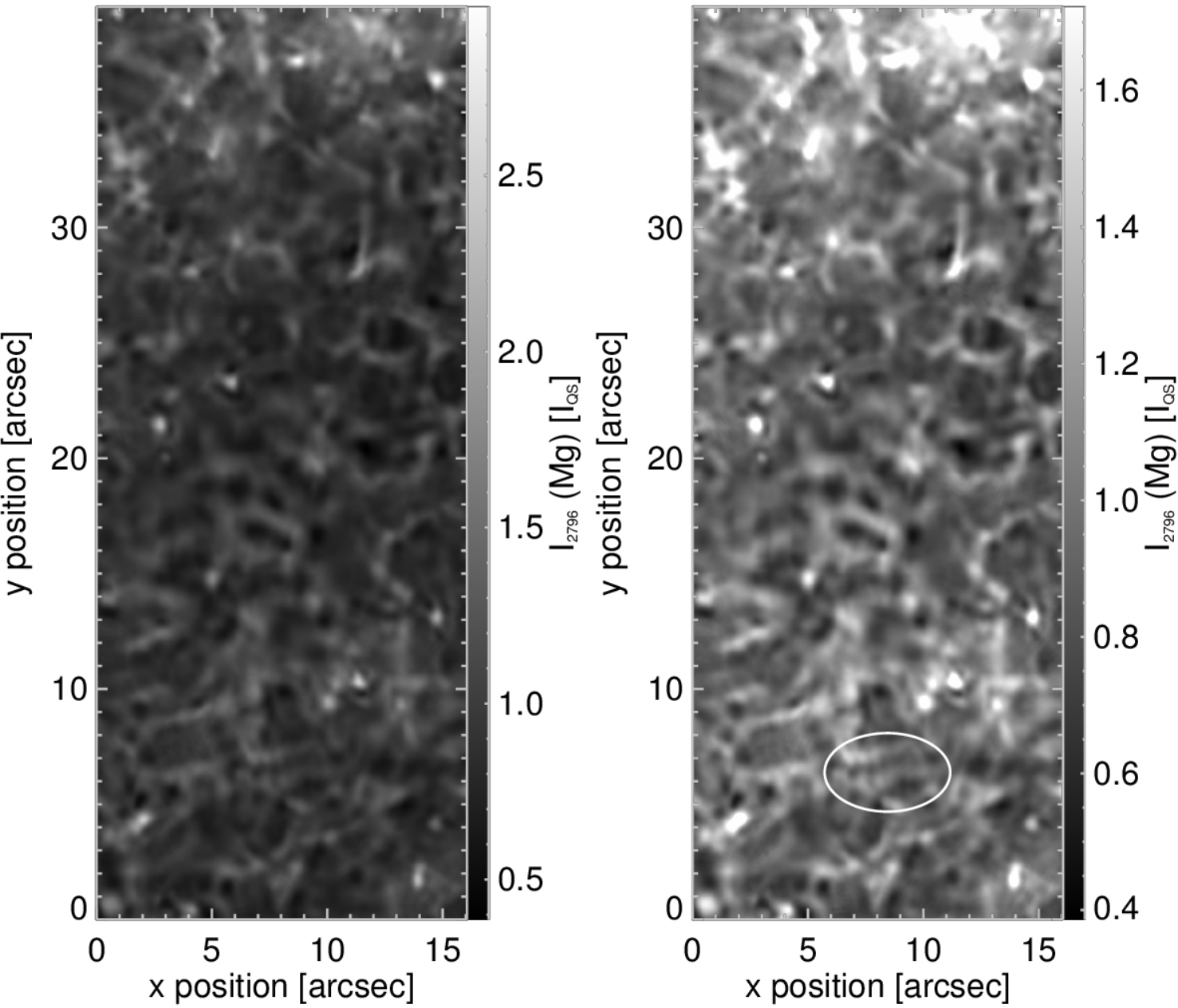}
   \caption{Broadband intensity image of a disk center quiet-Sun region observed in the Mg\,{\sc ii}\,k
   wavelength band on June 14 at 14:19~UT, normalized to the mean intensity level of the quiet Sun, $I_{\rm{QS}}$.
   The grey scale covers the full intensity range in the left panel, while it covers only three times the
   rms range of the image in the right panel, so that the brightest features are saturated. This view
   enhances the weaker structures. The oval marks a feature discussed in the text.}
   \label{Fig2}
   \end{figure}

   The flight altitude of on average $37$\,km provided seeing-free observations,
   nonetheless the pointing accuracy was changing due to varying shear wind strength, so that the CWS control
   loop could not be closed all the time. In total we got 105 focussed Magnesium images during the 2013
   flight from which we show one of the highest resolution ones obtained in a quiet part of the Sun, at
   solar disk center, in Figure~\ref{Fig2}. It looks qualitatively similar to high-resolution
   images recorded in Ca\,{\sc ii}\,H, e.g. during the first \sunrise{} flight \citep{Solanki2010,Jafarzadeh2013}:
   It is criss-crossed by brightenings roughly on the granular scale. In between are bright points present
   at various locations in the internetwork. These appear to be similar to Ca\,{\sc ii}\,H bright points
   \citep[see][and references therein]{Jafarzadeh2013}. A concentration of bright features is visible near
   the top of the frame in a strong network element. The resolution of the image is
   approximately 0\carcsec{}2. This value is deduced from both, measuring the sizes of the smallest
   visible structures and from the cut-off of the noise filter in the PD reconstruction, which is
   self-consistently computed from the data \citep{Loefdahl1994}.
   
   This resolution, although considerably higher than previous images taken in Mg\,{\sc ii} h or k, is
   below the diffraction limit. It is {\it a priori} unclear if the low resolution is mainly due to accumulated
   jitter during the $50$\,s exposure, or if it is caused by the evolution of the chromospheric
   structures during the exposure. Note that gondola-jitter was largest around local noon,
   when the telescope was pointing upward at a slanting angle, a less stable configuration than at times
   when the Sun was closer to the horizon and the telescope pointed more horizontally. 

   The rms intensity contrast of the displayed region is 24.2\%. This value is higher than the contrasts
   obtained by \cite{Hirzberger2010} in Ca\,{\sc ii}\,H (21-22\%). The comparison is not so clear-cut, however,
   due to the higher activity level during the 2013 flight, and possibly a difference in spatial resolution.
   In order to judge the effect of having an increased number and size of network features (possibly
   enhanced network) in the field of view of the Mg\,{\sc ii}\,k data, we additionally determined the rms
   contrast exclusively of clearly internetwork features, e.g. the region between 12\arcsec{} and 30\arcsec{}
   in the y direction of Fig.~\ref{Fig2}. This gave a value of 19.1\% which is at the lower end of the
   Ca\,{\sc ii}\,H intensity contrasts found by \cite{Hirzberger2010}, who did not explicitly exclude
   network features from their data. Therefore, disregarding the different filter widths for Mg\,{\sc ii}\,k
   and Ca\,{\sc ii}\,H observations, the resulting rms intensity contrasts are very similar.

   The grey scale in the right panel of Figure~\ref{Fig2} is limited to three times the rms range of the
   image which leads to a better visibility of reverse granulation (or possibly shock waves)
   that we find at the Mg wavelength. Note that in the lower part of the figure a similar structure
   is present as had first been noticed by \cite{Solanki2010} in Ca\,{\sc ii}\,H images
   obtained during the 2009 flight of \sunrise{}: there are often 2 parallel bright stripes, between which
   a dark lane is located (see the encircled example). Obviously the Mg\,{\sc ii}\,k line displays a similar
   structure. This structure may have to do with the width of the filter, since the
   monochromatic synthetic Mg\,{\sc ii} images computed by \cite{Leenaarts2013a,Leenaarts2013b} do not
   display this feature. 

   We detected 42 bright points (BPs) with the manual method used by \citet{Riethmueller2010} and determined
   their peak intensities. The mean BP peak intensity in Mg\,{\sc ii}\,k was found to be
   $(3.2 \pm 0.7)~I_{\rm{QS}}$, the highest value being $5.0~I_{\rm{QS}}$. These values are considerably
   higher than the contrasts found by \citet{Riethmueller2010} in the Ca\,{\sc ii}\,H core, with an average contrast
   of approximately $1.9$ and a maximum value of $3.4$. Note, however, that the Ca\,{\sc ii}\,H contrasts
   were obtained in the broad $1.8$\,\AA{} filter; as we shall see below the Mg\,{\sc ii}\,k images are
   more strongly reminiscent of the images recorded with the narrower $1.1$\,\AA{} Ca\,{\sc ii}\,H filter.
   The BPs in Mg\,{\sc ii} have a mean width of 0\carcsec{}55. This value is larger than the mean
   found in Ca\,{\sc ii} images by \citet{Keys2013} (0\carcsec{}35) and 0\carcsec{}2 by \citet{Jafarzadeh2013}.
   The latter authors restricted themselves to particularly small BPs (only those with diameters $< 0$\carcsec{}3
   were considered).

\begin{figure}
   \centering
   \includegraphics[width=\linewidth]{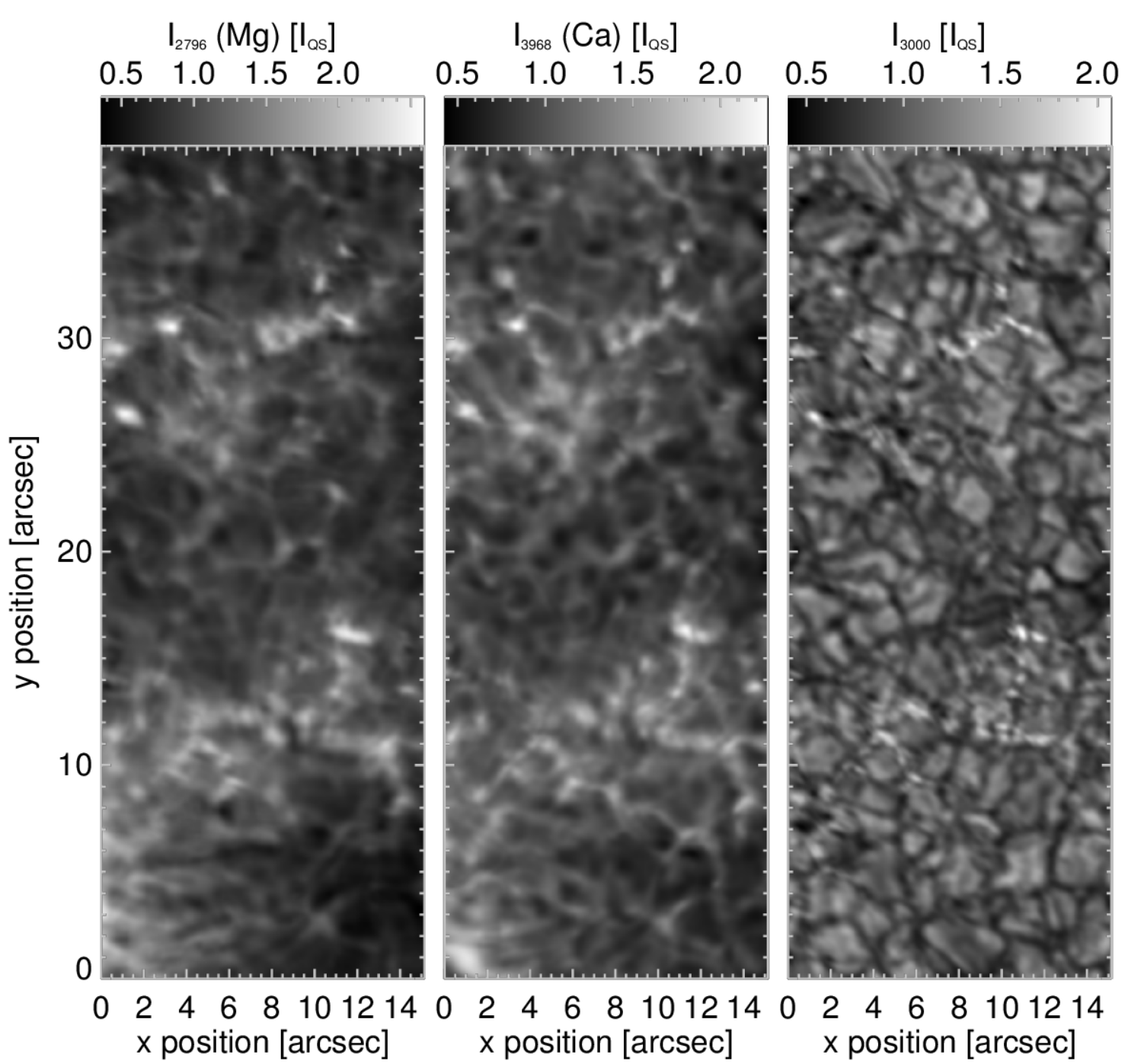}
   \caption{Intensity maps in the Mg\,{\sc ii}\,k, Ca\,{\sc ii}\,H, and $3000$\,\AA{} wavelength bands
   observed by SuFI on June 12 at 12:25~UT. The telescope pointed to a region at disk center with patches
   of quiet Sun alternating with bands of plage.}
   \label{Fig3}
   \end{figure}

   Nearly simultaneous SuFI observations at the wavelengths of Mg\,{\sc ii} $2796$\,\AA{}, the narrow
   ($1.1$\,\AA{}) band Ca\,{\sc ii} $3968$\,\AA{}, and $3000$\,\AA{} are displayed in Figure~\ref{Fig3}.
   The FOV exhibits, in addition to quiet granulation, two intrusions of plage around y=10\arcsec{}
   and y=30\arcsec{}. While the $3000$\,\AA{} wavelength is mainly formed in the deep photosphere and
   shows normal granulation in the quiet regions, the Mg\,{\sc ii} $2796$\,\AA{} and Ca\,{\sc ii}
   $3968$\,\AA{} wavelengths obviously look more chromospheric, i.e. the granulation pattern is reversed,
   although some of the bright structures may be signatures of shock waves.
   The contrast is higher and the bright features are more diffuse than in the photosphere.
   Furthermore, the bright features are also more diffuse in Mg\,{\sc ii}\,k than in Ca\,{\sc ii}\,H.
   In the lower part, below 10\arcsec{}, of the two chromospheric panels we find first stirrings of
   loop-like structures. These are clearly more pronounced in the Mg\,{\sc ii} line than in Ca\,{\sc ii}.  
   
   Since the solar scene is not totally quiet, the rms intensity contrast calculated over the entire
   Mg\,{\sc ii} image, 29.8\%, is distinctly higher than in Figure~\ref{Fig2}. For the Ca\,{\sc ii}
   $3968$\,\AA{} and $3000$\,\AA{} image we find 22.5\% and 20.4\%, respectively. Interestingly, these last
   two values are not higher than what \cite{Hirzberger2010} found in the quiet Sun. For $3000$\,\AA{} this
   is acceptable since the density of BPs is relatively low when averaged over the whole FOV. However,
   for Ca\,{\sc ii} it is surprising since the wavelength band of this image is narrower than at the time of
   the 2009 flight. One implication may be that either the features become more diffuse at the greater
   heights sampled by Ca\,{\sc ii} in the  narrower filter, or that this particular image is less sharp
   than some of the sharpest ones from the first flight of \sunrise{}.  


\begin{figure}
   \centering
   \includegraphics[width=\linewidth]{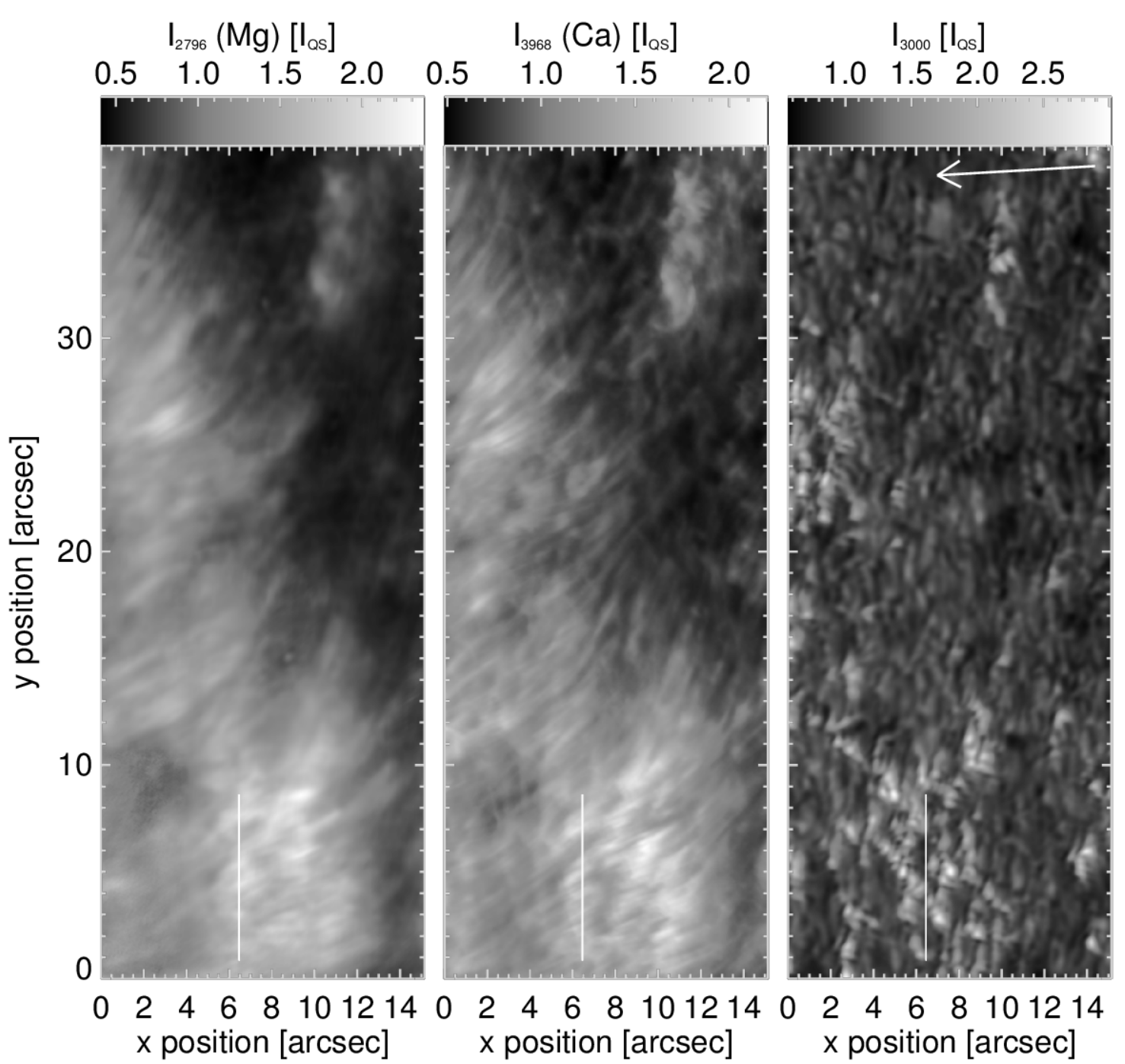}
   \caption{Same as Fig.~\ref{Fig3} but observed on June 16 at 13:13~UT while the telescope pointed to a
   plage region between the two sunspots of AR11768 ($\mu = 0.41$). The white arrow in the top right corner
   of the image points towards the disk center. The white lines in the bottom part mark
   the cuts shown in Fig.~\ref{Fig5}.}
   \label{Fig4}
   \end{figure}

\begin{figure}
   \centering
   \includegraphics[width=\linewidth]{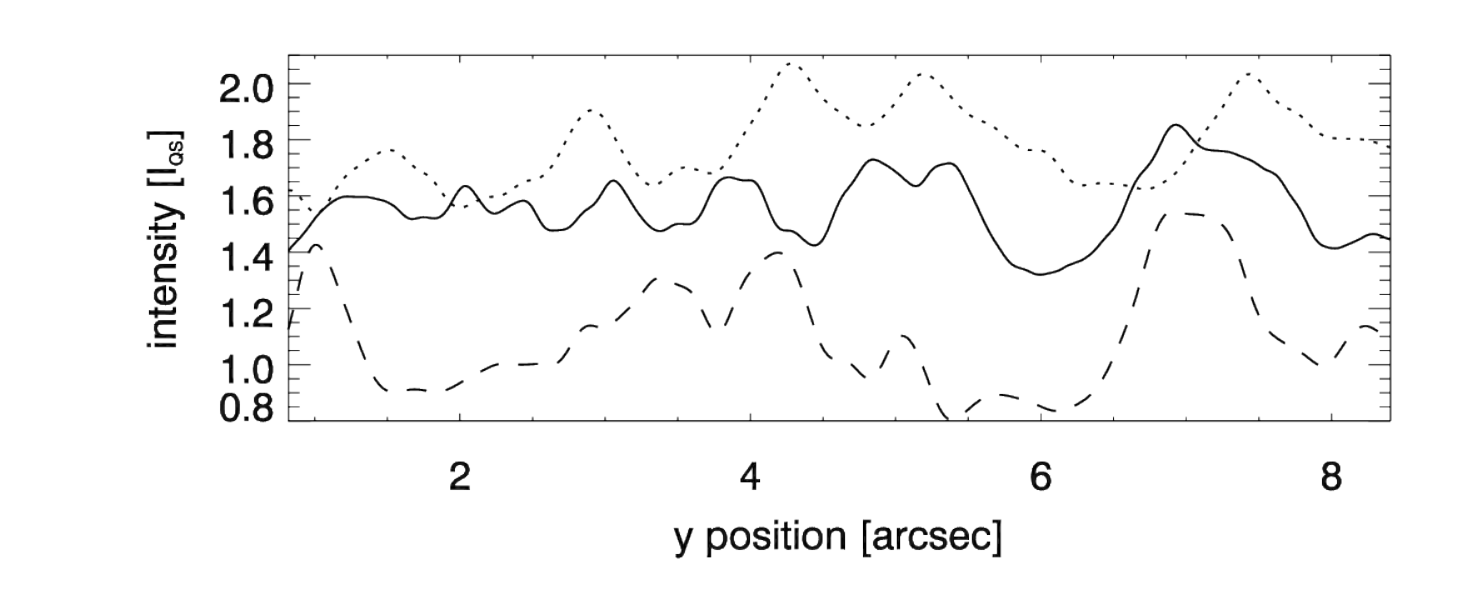}
   \caption{Intensity cuts at the positions of the white lines in Fig.~\ref{Fig4}:
   Mg\,{\sc ii} - dotted line, Ca\,{\sc ii} - solid line, $3000$\,\AA{} - dashed line.}
   \label{Fig5}
   \end{figure}

   Fibrils become clearly visible when observed nearer to the limb in regions of high
   magnetic activity. This is displayed in Fig.~\ref{Fig4}, in which a part of AR11768 is depicted,
   observed at $\mu = 0.41$. The field-of-view was located between the main sunspots
   of the active region. While the photospheric $3000$\,\AA{} image shows numerous faculae that are typical
   of active regions at higher heliocentric angles, the two chromospheric images clearly reveal a forest of
   bright chromospheric fibrils. The fibrils in this figure are more diffuse in Mg\,{\sc ii}
   than in Ca\,{\sc ii} as demonstrated by the intensity cuts across fibrils in Fig.~\ref{Fig5}.
   Note that the structure along the fibrils (e.g. due to finite length of the fibrils) looks
   rather similar in both the Mg\,{\sc ii} and Ca\,{\sc ii}.

\section{Discussion and conclusions}

   We have presented the first high-resolution images of the solar atmosphere taken in the
   Mg\,{\sc ii}\,k line. These were obtained with the SuFI instrument on the \sunrise{}
   balloon-borne solar observatory during its second science flight that took place on 12-17
   June 2013. The atmospheric absorption by stratospheric ozone is still very strong, even at
   altitudes above $37$\,km, so that long integration times were needed, and only images with relatively
   low S/N were obtained.  

   The images display a pattern that may be due to reversed granulation or shock waves in the quiet-Sun
   internetwork, very similar to interference-filter images taken in the core of Ca\,{\sc ii}\,H.
   Figure~3 of \cite{Leenaarts2013b}, who computed the Mg\,{\sc ii} h\&k lines in a snapshot from a
   radiation MHD simulation carried out with the Bifrost code \citep{Gudiksen2011}, displays such a
   pattern at the Mg\,{\sc ii}\,k2v and k2r wavelengths, but not in the Mg\,{\sc ii}\,k3 image.
   This suggests that the \sunrise{}/SuFI Mg\,{\sc ii}\,k filter samples the k2 peaks much more than k3,
   which is not surprising given the greater strength of the peaks and their larger spectral coverage. 

   The rms contrast in Mg\,{\sc ii}\,k of these very quiet regions is 19.1\%, obtained after masking
   out network areas in quiet Sun images. This is very close to the rms contrast of the quiet Sun obtained by
   \cite{Hirzberger2010} in Ca\,{\sc ii}\,H, also with \sunrise{}/SuFI during its first science flight in 2009. 

   At disk center Mg\,{\sc ii}\,k also shows bright points at the same locations at which such features
   are seen in the photospheric $3000$\,\AA{} line and in the core of Ca\,{\sc ii}\,H. These  BPs are considerably
   brighter in Mg\,{\sc ii} than at other wavelengths. Mg\,{\sc ii}\,k provides a mean contrast of
   $3.2~I_{\rm QS}$, which is considerably higher than the values of $1.9~I_{\rm QS}$ and $2.3~I_{\rm QS}$
   found by \citet{Riethmueller2010} in Ca\,{\sc ii}\,H and at $2140$\,\AA{}, respectively, the two wavelengths
   displaying the largest BP contrasts during the first \sunrise{} flight. This makes the
   Mg\,{\sc ii}\,k line very promising for identifying and tracking small-scale magnetic features,
   although the structures may be more diffuse than at photospheric heights (see below). Note, however,
   that the computations of \cite{Leenaarts2013b}, do not display any particular brightening at the
   footpoints of the emerging bipole in k2v and k2r. One of the footpoints is bright in k3, but the
   picture is quite different from the observations. We speculate that the difference has to do with
   the width of the SuFI filter, which samples also considerable portions of the inner
   wings of the Mg\,{\sc ii}\,k line. These are formed mainly in the upper photosphere, where kilo-Gauss
   magnetic features appear as bright point-like structures.  

   The Mg images also display fibrils emanating from plage regions. The presence of such fibrils had
   been predicted by \citet{Leenaarts2013a,Leenaarts2013b}. Not unexpectedly, the density of fibrils
   increases very strongly from k2v and k2r to k3. The density of fibrils in our images is closer to
   that in k2v and k2r, although only a very qualitative comparison can be carried out, since we find
   that the density of fibrils depends very strongly on the magnetic activity level (i.e. on how
   strong the plage is). This is in line with the behaviour of Ca\,{\sc ii}\,H. Also, possibly the
   fibril density is larger closer to the limb (this will have to be studied
   in future investigations). At disk center the fibrils seen in Mg\,{\sc ii} are more pronounced than
   in Ca\,{\sc ii}\,H \citep[see, e.g.][for a study of the fibrils seen in Ca\,{\sc ii}\,H]{Pietarila2009}. 

   Features in Mg\,{\sc ii}\,k are more diffuse than in Ca\,{\sc ii}\,H. E.g., the BPs
   have an average size of 0\carcsec{}55, which is significantly larger than in Ca\,{\sc ii}
   \citep{Keys2013,Jafarzadeh2013}. We expect the following possible reasons for this:  because
   a) Mg\,{\sc ii} is formed higher in the chromosphere than Ca\,{\sc ii} \citep{Leenaarts2013a} or
   b) because of the much longer exposure time of Mg\,{\sc ii} images, which allows the features to
   evolve during the exposure, or c) due to residual jitter not completely compensated by the image
   stabilization system. A study of these causes as well as a more detailed and
   quantitative comparison between Mg\,{\sc ii} and Ca\,{\sc ii} will be the topic of a separate paper
   (Danilovic et al. in preparation). 

   In summary, a very first analysis of some of the images recorded by \sunrise{} during its 2013 flight
   has shown that observations of the Mg\,{\sc ii}\,k line with a $4.8$\,\AA{} interference filter provide
   a qualitatively similar picture as obtained from the narrowest interference filter observations of
   the Ca\,{\sc ii}\,H line core. However, there are considerable quantitative differences, partly due to
   the different formation of the lines, partly due to observing constraints. It will be interesting to
   continue such an analysis using data from both \sunrise{} and the IRIS mission.   

   \begin{acknowledgements}
   The German contribution to the \sunrise{} reflight has been funded by the Max Planck Foundation, the
   Innovations Fund of the President of the Max Planck Society (MPG) and private donors, which is gratefully
   acknowledged. The Spanish contribution has been funded by the Spanish MICINN under the project
   AYA2012-39636-C06. The HAO contribution has been partly funded through NASA grant number NNX13AE95G.
   This work has been partly supported by the WCU grant (No. R31-10016) funded by the Korean Ministry of
   Education, Science \& Technology.
   \end{acknowledgements}


\end{document}